\documentclass[aps,prl,showpacs,twocolumn]{revtex4}
\usepackage{graphicx}
\usepackage{dcolumn}
\usepackage{multirow}
\begin{document} 

\title{Atomic and electronic structure of the $\sqrt{3}$$\times$$\sqrt{3}$ silicene phase on Ag(111): density-functional calculations}
\author{Se Gab Kwon}
\author{Myung Ho Kang}
\email{kang@postech.ac.kr}
\affiliation{Department of Physics, Pohang University of Science and Technology, Pohang 790-784, Korea}
\date{\today}

\begin{abstract}
Density-functional theory calculations are used to verify the atomic structure of the $\sqrt{3}$$\times$$\sqrt{3}$ silicene phase grown on the Ag(111) surface.
Recent experimental studies strongly suggested that the previous double-layer silicene model should be replaced with a Ag-mixed double-layer model resembling the top layers of the Ag/Si(111)-($\sqrt{3}$$\times$$\sqrt{3}$) surface.
In our calculations, the Ag-mixed double-layer model is indeed energetically favored over the double-layer silicene model and well reproduces the reported scanning-tunneling microscopy and spectroscopy data.
Especially, the structural origin of the experimental band structure is clarified as the top Ag-Si mixed layer, unlike the experimental interpretations as either a silicene layer or the Ag(111) substrate.  
\end{abstract}

\pacs{68.43.Bc, 71.15.Mb, 81.07.Vb}
\maketitle

\section{I. INTRODUCTION} 


Silicene, a monolayer of Si atoms arranged in a honeycomb lattice, has attracted great attention as a new Dirac-type electron system similar to graphene \cite{caha09,liuu11,ezaw12,ezaw13,pann14}.
One practical way of preparing silicene is to grow it on metallic or half-metallic substrates such as ZrB$_{2}$(0001) \cite{fleu12}, Ir(111) \cite{meng13}, and Ag(111) \cite{linn13}.
Recent studies have been focused on the silicene fabricated on Ag(111) by thermal deposition of Si, exhibiting a variety of structural phases including 3$\times$3 \cite{vogt12,majz13,liuu14}, $\sqrt{3}$$\times$$\sqrt{3}$ \cite{chen12,chen13,chen13'}, and $\sqrt{7}$$\times$$\sqrt{7}$ \cite{linn12,john14,tcha14} with respect to the silicene unit cell, depending on the growth temperature and Si coverage.


Particularly interesting is the $\sqrt{3}$$\times$$\sqrt{3}$ silicene phase grown on Ag(111).
This phase has been highlighted with exotic physical properties such as Dirac fermion charge carriers \cite{chen12}, temperature-induced structural transition \cite{chen13}, and superconductivity \cite{chen13'}, but its atomic structure remains controversial.
Chen and co-workers \cite{chen12,chen13} first proposed a single-layer silicene model, but it was questioned by scanning tunneling microscopy (STM) \cite{rest13,pado13,araf13} and atomic force microscopy (AFM) \cite{rest13} studies, in which the measured topographic heights imply a double-layer thickness of the silicene.
A quantitative double-layer silicene model (see Fig. 1) was soon provided by Guo and Oshiyama by means of density functional theory (DFT) calculations \cite{guo14}. 


This double-layer silicene model, however, was not supported by a low-energy electron diffraction (LEED) study of Shirai and co-workers \cite{shir14}: Their LEED data could not be explained by the double-layer silicene model, but instead were similar to that of the Ag/Si(111)-($\sqrt{3}$$\times$$\sqrt{3}$) surface.
The Ag/Si(111)-($\sqrt{3}$$\times$$\sqrt{3}$) surface is well known for the Ag-induced surface reconstruction, referred to as the inequivalent-triangle model \cite{aiza99}, in which Ag atoms effectively replace the surface Si atoms with dangling bonds.  
Therefore, Shirai and co-workers \cite{shir14} suggested that the top part of the silicene double layer should be replaced with Ag atoms to resemble the adsorption structure of the Ag/Si(111)-($\sqrt{3}$$\times$$\sqrt{3}$) surface, as schematically shown in Fig. 1(b). 
Interestingly, Mannix and co-workers \cite{mann14} also independently reached the same mixed double-layer (MDL) picture from the similar STM images between the $\sqrt{3}$$\times$$\sqrt{3}$ silicene phase on Ag(111) and the Ag/Si(111)-($\sqrt{3}$$\times$$\sqrt{3}$) surface.
The energetical, microscopic and spectroscopic tests of the experimental MDL model, however, is yet to be done. 


\begin{figure}
\centering{\includegraphics[width=8cm]{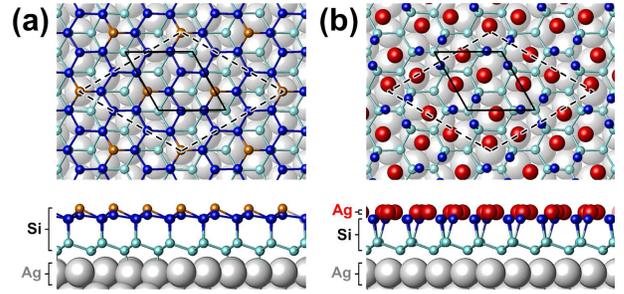}}
\caption{\label{fig1}
(Color online) 
Structural models for the $\sqrt{3}$$\times$$\sqrt{3}$ silicene phase on Ag(111).
(a) Double-layer silicene model.
(b) Mixed double-layer (MDL) model.
Small balls represent Si atoms and large (medium) balls the substrate (top-layer) Ag atoms.
Small and large boxes represent the silicene-($\sqrt{3}$$\times$$\sqrt{3}$) and Ag(111)-(4$\times$4) unit cells, respectively.}

\end{figure}


In this paper, we use DFT calculations to examine the MDL model for the $\sqrt{3}$$\times$$\sqrt{3}$ silicene phase grown on Ag(111). 
The MDL model is found to form a stable adsorption structure on the Ag(111) substrate, which is energetically favored over the earlier double-layer silicene model.
The MDL model is also sound both microscopically and spectroscopically: Its electronic structure accounts well for the measured STM images and band structures.

\section{II. METHOD} 

We perform DFT calculations using the Vienna $ab$ $initio$ simulation package \cite{kres96} within the Perdew-Burke-Ernzerhof generalized gradient approximation \cite{perd96} and the projector augmented wave method \cite{bloc94,kres99}.
The Ag(111) surface is modeled by a periodic slab geometry with five atomic layers and a vacuum spacing of about 19 {\AA}.
Si and Ag atoms are incorporated on the top of the Ag(111)-(4$\times$4) slab unit cell, a supercell  commensurating with the silicene-($\sqrt{3}$$\times$$\sqrt{3}$) lattice \cite{guo15}.
The calculated value 2.938 {\AA} is used as the bulk Ag-Ag bond length, which is in good agreement with the experimental value 2.892 {\AA} \cite{kitt05}.
We expand the electronic wave functions in a plane-wave basis with an energy cutoff of 250 eV.
A (6$\times$6$\times$1) $k$-point mesh is used for the (4$\times$4) Brillouin-zone integrations.
All atoms but the bottom two Ag layers are relaxed until the residual force components are within 0.02 eV/{\AA}.
The used slab and vacuum thickness, plane-wave energy, and $k$-point mesh were found to produce sufficiently converged results for the MDL model: the formation energy and the interatomic distances converge well within 0.05 eV and 0.02 \AA, respectively.
We compare the energetics between different models by estimating the relative formation energy, defined by $\Delta$$E=E_{\rm 1}$$-$$E_{\rm 0}$$-$$\Delta$$n_{\rm Ag}$$\mu_{\rm Ag}$$-$$\Delta$$n_{\rm Si}$$\mu_{\rm Si}$, where $E_{\rm 1}$ ($E_{\rm 0}$) is the total energy of a particular model (a reference model), $\Delta$$n$ is the number difference of the specified atoms relative to the reference model, and $\mu$ is the calculated bulk chemical potential for the specified atom.

\section{III. RESULTS} 


\begin{figure} [b]
\centering{\includegraphics[width=8cm]{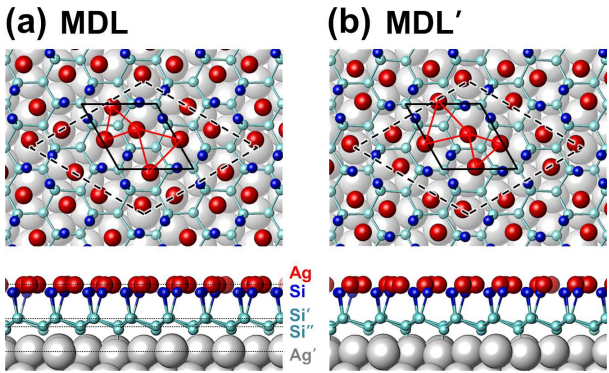}}
\caption{\label{fig2}
(Color online) 
Two optimized MDL structures.
}
\end{figure}


Figure 2 shows two energetically-degenerated MDL structures under consideration, which were well established in the Ag/Si(111)-($\sqrt{3}$$\times$$\sqrt{3}$) system \cite{aiza99}. 
The MDL structure is characterized by two inequivalent Ag triangles, and the MDL$'$ structure is distinguishable with the mirror-symmetric conversion of the two Ag triangles. 
The MDL and MDL$'$ structures can be transformed to each other by relatively small displacements of the top Ag ($\sim$0.5 {\AA}) and Si ($\sim$0.3 {\AA}) atoms. 
We calculated the corresponding energy barrier by using the nudged elastic-band method \cite{jons98} that allows an effective search for the lowest-energy path for a multi-atomic structural transition.
The result is as low as 66 meV per $\sqrt{3}$$\times$$\sqrt{3}$ silicene unit cell.
This low energy barrier and the resulting thermal fluctuations will be discussed below in accounting for the STM feature that the $\sqrt{3}$$\times$$\sqrt{3}$ silicene phase on Ag(111) appears differently at above 40 K from two distinct mirror-symmetric images at lower temperatures \cite{chen13}.


\begin{table}
\caption{\label{tab:table1}
Structural details of the MDL model.
$d$ (\AA) represents the bond length of the Ag or Si trimer in the top layer.
$h$ (\AA) represents the average interlayer spacing between specified atomic layers.}
\begin{ruledtabular}
\renewcommand{\arraystretch}{1.5}
\begin{tabular}{cccccc}
$d_{\rm Ag-Ag}$ & $d_{\rm Si-Si}$ & $h_{\rm Ag-Si}$ & $h_{\rm Si-Si'}$ & $h_{\rm Si'-Si''}$ & $h_{\rm Si''-Ag'}$\\
\hline
3.01 & 2.54 & 0.69 & 2.32 & 0.80 & 2.26 \\
\end{tabular}
\end{ruledtabular}
\end{table}


Table I summarizes the structural details of the MDL model shown in Fig. 2(a).
The Ag-Ag (3.01 \AA) and Si-Si (2.54 \AA) bond lengths and the Ag-Si (0.69 \AA) and Si-Si$'$ (2.32 \AA) layer spacings compare well with those (3.00, 2.58, 0.72, and 2.29 \AA, respectively) of the inequivalent-triangle  Ag/Si(111)-($\sqrt{3}$$\times$$\sqrt{3}$) surface \cite{aiza99}.
The mixed double layer is separated by 2.26 {\AA} from the Ag(111) substrate.
The resulting Si-Ag bond lengths of about 2.56 {\AA} implies a covalent bonding between Si and Ag atoms in view that the covalent radii of Si and Ag are 1.11 {\AA} and 1.45 {\AA} \cite{cord08}, respectively, as was already argued by Guo and Oshiyama in their DFT study of the double-layer silicene model \cite{guo14}.


The MDL model is more stable in energy than the double-layer silicene model of Guo and Oshiyama \cite{guo14}.
Its lower formation energy by 1.02 eV per $\sqrt{3}$$\times$$\sqrt{3}$ silicene unit cell is possibly attributed to the removal of Si dangling bonds by Ag substitution for the top-layer Si atoms.
On the other hand, the MDL model has 0.66 eV higher formation energy than the well-established 3$\times$3 single-layer silicene phase \cite{vogt12}, supporting the experimental finding that, as increasing the Si coverage, the $\sqrt{3}$$\times$$\sqrt{3}$ silicene phase appears after covered with the 3$\times$3 phase \cite{vogt14}. 


The MDL model accounts well for the apparent height difference in STM and AFM topographs between the $\sqrt{3}$$\times$$\sqrt{3}$ and the 3$\times$3 silicene phases on Ag(111) \cite{rest13,pado13,araf13}.
In our calculations, the topmost layer of the present MDL model is 3.14 {\AA} higher than that of the 3$\times$3 single-layer silicene structure \cite{vogt12}, which compares well with the AFM measurement of 3.0 {\AA} \cite{rest13} and the STM measurement of 2.2 {\AA} \cite{pado13,araf13}.
Somewhat smaller STM result may possibly reflect different surface electronic structures of the two phases.


\begin{figure}
\centering{\includegraphics[width=8cm]{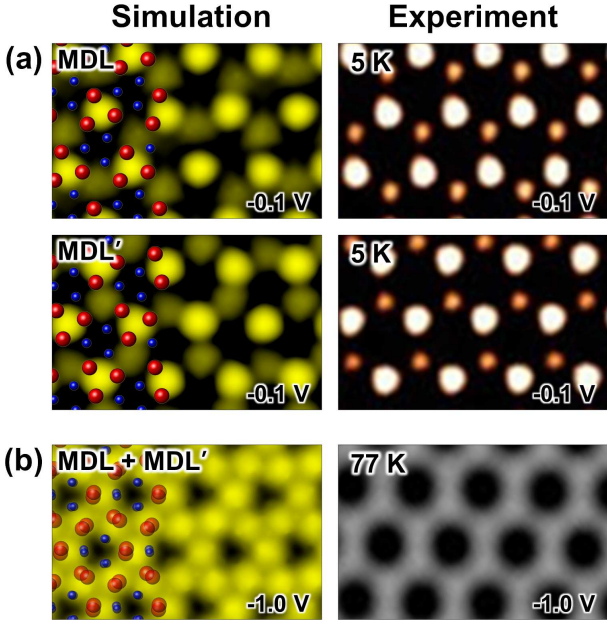}}
\caption{\label{fig3}
(Color online)
Simulated STM images in comparison with experimental data.
(a) Simulations for the MDL and MDL$'$ structures.
The images represent the surface of constant density with $\rho$=1$\times$10$^{-5}$ e/\AA$^{3}$ taken at the bias voltage $-$0.1 V.
The STM images obtained at 5 K were taken from Ref. \cite{chen13}.
(b) Simulated image obtained by superimposing the images of the MDL and MDL$'$ structures.
The image represents the surface of constant density with $\rho$=1$\times$10$^{-4}$ e/\AA$^{3}$ taken at the bias voltage $-$1.0 V.
The STM image obtained at 77 K was taken from Ref. \cite{chen13}.
}
\end{figure}


Figure 3(a) shows the simulated STM images of the two mirror-symmetric MDL structures shown in Fig. 2.
Both images feature two superposed hexagonal arrays: one with bright spots and the other with weaker spots, representing the centers of small and large Ag triangles, respectively.
Here, the top-layer Si triangles appear dark in both images.
It is experimentally known that the $\sqrt{3}$$\times$$\sqrt{3}$ silicene phase shows two mirror-symmetric configurations separated by narrow domain boundaries at low temperatures below 40 K \cite{chen13}.
The STM images measured at 5 K, representing two different domains, indeed compare well with our simulations.
At elevated temperatures, however, thermal fluctuations may take place between the two mirror-symmetric (MDL and MDL$'$) structures.
Figure 3(b) shows the superposition of the two mirror-symmetric images, appearing as a prominent honeycomb array, which compares well with the STM image taken at 77 K \cite{chen13}.
The experimental honeycomb image appears more symmetric than the simulation possibly due to the rounding effect of thermal fluctuations.”


\begin{figure*}
\centering{\includegraphics[width=16cm]{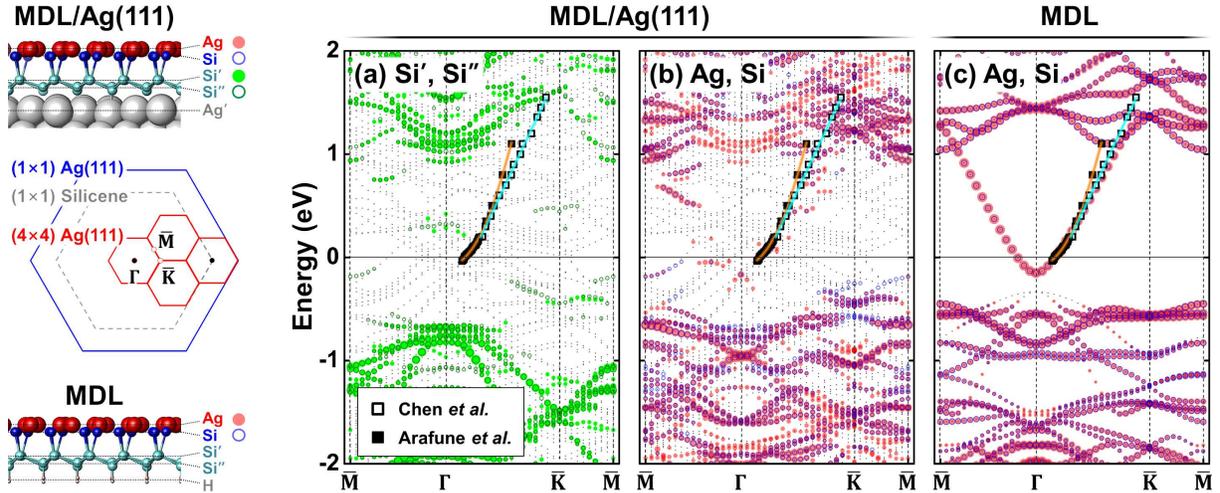}}
\caption{\label{fig4}
(Color online)
Band structure of the MDL model.
(a) Filled (open) circles represent the states derived from the Si$'$ (Si$''$) layer.
(b) Filled (open) circles represent the states derived from the Ag (Si) atoms in the mixed layer.
Their size is proportional to the amount of charge localized in the corresponding atomic layers.
Open and filled squares represent the experimental data of Chen $et$ $al$. \cite{chen13} and Arafune $et$ $al$. \cite{araf13}, respectively.
(c) Band structure of the freestanding MDL structure. 
Filled (open) circles represent the states derived from the Ag (Si) atoms in the mixed layer.
With the Ag(111) substrate removed, the states coming from the surface Ag atoms become prominent and compare well with the experiments.
}
\end{figure*}


Figure 4 shows the calculated band structure of the MDL model in comparison with scanning tunneling spectroscopy (STS) measurements \cite{chen13,araf13}.
While Chen and co-workers \cite{chen13} reported an almost linear band (denoted by open squares) and attributed its origin to the silicene layer, Arafune and co-workers \cite{araf13} claimed that their STS band (denoted by filled squares) must be described by a parabolic function and originates from the Ag(111) substrate.
In our layer-resolved band analysis, however, the experimental bands could be related to neither the Si$'$ and Si$''$ layers of the mixed double layer nor the Ag(111) substrate, as seen in Fig. 4(a).
They instead can be related to the Ag and Si mixed layer although the calculated band character appears rather weak possibly due to the resonance mixing with the broadly-spread bulk states steming from the Ag(111) substrate. 
In order to clarify the band picture of the Ag and Si mixed layer, we examined the band structure of the isolated MDL structure, the bottom of which is now terminated by H atoms instead of the Ag(111) substrate.
Here, as shown in Fig. 4(c), the Ag and Si mixed layer produces a prominent surface band near the Fermi level, exhibiting a nearly parabolic shape.
A parabolic fitting to the bottom part (below 0.2 eV) results in an effective mass of 0.11$m_{e}$ ($m_{e}$ is the mass of free electron), comparable to the experimental value 0.14$m_{e}$ reported by Arafune and co-workers \cite{araf13}.
In the range of 0.2--1.0 eV, the band shows an almost linear dispersion, and the calculated velocity of 0.84$\times$10$^{6}$ m/s also compares well with the experimental value 0.97$\times$10$^{6}$ m/s reported by Chen and co-workers \cite{chen13}.
Therefore, the present MDL model not only reproduces the experimental band structure but also corrects the previous misinterpretations of the band origin: It is not from either a silicene layer or the Ag(111) substrate, but from the top Ag-Si mixed layer.  


In light of the present confirmation of the MDL model, it is noticeable that a recent STM study refuted the possibility of Ag segregation to the top silicene layer. 
In their STM study of $\sqrt{3}$$\times$$\sqrt{3}$ multi-layer silicene films grown on Ag(111) \cite{chen15}, Chen and co-workers applied bias pulses to the tip on the surface, thereby removing a part of the surface layers beneath the tip at liquid nitrogen temperature.
Interestingly, the underneath exposed layer (supposed to be a pure silicene layer) also showed a similar $\sqrt{3}$$\times$$\sqrt{3}$ STM image to the top layer, and they inferred from the similar STM images that the top layer is also a pure silicene layer. 
It should be mentioned, however, that the similar STM images do not necessarily guarantee the identical atomic structures. 
In fact, the double-layer silicene model by Guo and Oshiyama \cite{guo14} and the present MDL model are equally compatible with the experimental $\sqrt{3}$$\times$$\sqrt{3}$ honeycomb image \cite{chen13}. 
So, the top mixed layer and deeper pure silicene layers could also be compatible with the reported STM observations of Chen and co-workers \cite{chen15}. 
Only the comparison of the STM images is not yet sufficient for a definite chemical and structural assignment.

\section{IV. SUMMARY} 

The present DFT calculations confirmed the experimentally-proposed mixed double-layer model for the $\sqrt{3}$$\times$$\sqrt{3}$ silicene phase grown on Ag(111) by demonstrating that it is energetically sound and well reproduces the reported STM images and STS band structures. 

\section{ACKNOWLEDGMENTS} 

This work was supported by the National Research Foundation of Korea (Grant No. 2011-0008907).


\newcommand{\PR} [3]{Phys.\ Rev.\ {\bf #1}, #2 (#3)}
\newcommand{\PRL}[3]{Phys.\ Rev.\ Lett.\ {\bf #1}, #2 (#3)}
\newcommand{\PRB}[3]{Phys.\ Rev.\ B\ {\bf #1}, #2 (#3)}
\newcommand{\PST}[3]{Phys.\ Scr.\ T\ {\bf #1}, #2 (#3)}
\newcommand{\PML}[3]{Phil.\ Mag.\ Lett.\ {\bf #1}, #2 (#3)}
\newcommand{\SCI}[3]{Science {\bf #1}, #2 (#3)}
\newcommand{\SSA}[3]{Surf.\ Sci.\ {\bf #1}, #2 (#3)}
\newcommand{\SSCO}[3]{Solid\ State\ Comm.\ {\bf #1}, #2 (#3)}
\newcommand{\SSL}[3]{Surf.\ Sci.\ Lett.\ {\bf #1}, #2 (#3)}
\newcommand{\SRL}[3]{Surf.\ Rev.\ Lett.\ {\bf #1}, #2 (#3)}
\newcommand{\NAT}[3]{Nature\ Phys.\ {\bf #1}, #2 (#3)}
\newcommand{\JP}[3]{J.\ Phys.\ {\bf #1}, #2 (#3)}
\newcommand{\JACS}[3]{J.\ Am.\ Chem.\ Soc.\ {\bf #1}, #2 (#3)}
\newcommand{\JAP}[3]{J.\ Appl.\ Phys.\ {\bf #1}, #2 (#3)}
\newcommand{\JCP}[3]{J.\ Chem.\ Phys.\ {\bf #1}, #2 (#3)}
\newcommand{\JPCS}[3]{J.\ Phys.\ Chem.\ Solids.\ {\bf #1}, #2 (#3)}
\newcommand{\JVSA}[3]{J.\ Vac.\ Sci.\ Technol.\ A\ {\bf #1}, #2 (#3)}
\newcommand{\JVSB}[3]{J.\ Vac.\ Sci.\ Technol.\ B\ {\bf #1}, #2 (#3)}
\newcommand{\JJAP}[3]{Jpn.\ J.\ Appl.\ Phys.\ {\bf #1}, #2 (#3)}
\newcommand{\ASS}[3]{Appl.\ Surf.\ Sci.\ {\bf #1}, #2 (#3)}
\newcommand{\APL}[3]{Appl.\ Phys.\ Lett.\ {\bf #1}, #2 (#3)}
\newcommand{\CPL}[3]{Chem.\ Phys.\ Lett.\ {\bf #1}, #2 (#3)}
\newcommand{\LTP}[3]{Low\ Temp.\ Phys.\ {\bf #1}, #2 (#3)}
\newcommand{\TSF}[3]{Thin\ Solid\ Filims\ {\bf #1}, #2 (#3)}
\newcommand{\VAC}[3]{Vacuum\ {\bf #1}, #2 (#3)}
\newcommand{\JPCC}[3]{J.\ Phys.\ Chem.\ C\ {\bf #1}, #2 (#3)}
\newcommand{\RPP}[3]{Rep.\ Prog.\ Phys.\ {\bf #1}, #2 (#3)}
\newcommand{\JPCM}[3]{J.\ Phys.:\ Condens.\ Matter.\ {\bf #1}, #2 (#3)}
\newcommand{\CJCP}[3]{Chin.\ J.\ Chem.\ Phys.\ {\bf #1}, #2 (#3)}
\newcommand{\IJMPB}[3]{Int.\ J.\ Mod.\ Phys.\ B\ {\bf #1}, #2 (#3)}
\newcommand{\RMP}[3]{Rev.\ Mod.\ Phys.\ {\bf #1}, #2 (#3)}
\newcommand{\NNT}[3]{Nanotechnology\ {\bf #1}, #2 (#3)}
\newcommand{\JPD}[3]{J.\ Phys.\ D\ {\bf #1}, #2 (#3)}
\newcommand{\JPC}[3]{J.\ Phys.\ C\ {\bf #1}, #2 (#3)}
\newcommand{\PSSC}[3]{Phys.\ Status\ Solidi\ C\ {\bf #1}, #2 (#3)}
\newcommand{\EL}[3]{Europhys.\ Lett.\ {\bf #1}, #2 (#3)}
\newcommand{\JERP}[3]{J.\ Electron Spectrosc.\ Rel.\ Phen.\ {\bf #1}, #2 (#3)}
\newcommand{\NL}[3]{Nano\ Lett.\ {\bf #1}, #2 (#3)}
\newcommand{\AM}[3]{Adv.\ Mater.\ {\bf #1}, #2 (#3)}
\newcommand{\NJP}[3]{New\ J.\ Phys.\ {\bf #1}, #2 (#3)}
\newcommand{\APE}[3]{Appl.\ Phys.\ Express.\ {\bf #1}, #2 (#3)}
\newcommand{\SR}[3]{Sci.\ Rep.\ {\bf #1}, #2 (#3)}
\newcommand{\ACS}[3]{ACS\ Nano\ {\bf #1}, #2 (#3)}
\newcommand{\AFM}[3]{Adv.\ Funct.\ Mattert.\ {\bf #1}, #2 (#3)}
\newcommand{\DT}[3]{Dalton\ Trans.\ {\bf #1}, #2 (#3)}


\begin{thebibliography}{}
\bibitem{caha09} S. Cahangirov, M. Topsakal, E. Akt{\"{u}}rk, H. {\c{S}}ahin, and S. Ciraci, Two- and one-dimensional honeycomb structures of silicon and germanium, \PRL{102}{236804}{2009}.
\bibitem{liuu11} C.-C. Liu, W. Feng, and Y. Yao, Quantum spin hall effect in silicene and two-dimensional germanium, \PRL{107}{076802}{2011}.
\bibitem{ezaw12} M. Ezawa, Valley-polarized metals and quantum anomalous hall effect in silicene, \PRL{109}{055502}{2012}.
\bibitem{ezaw13} M. Ezawa, Photoinduced topological phase transition and a single dirac-cone state in silicene, \PRL{110}{026603}{2013}.
\bibitem{pann14} H. Pan, Z. Li, C.-C. Liu, G. Zhu, Z. Qiao, and Y. Yao, Valley-polarized quantum anomalous hall effect in silicene, \PRL{112}{106802}{2014}.
\bibitem{fleu12} A. Fleurence, R. Friedlein, T. Ozaki, H. Kawai, Y. Wang, and Y. Yamada-Takamura, Experimental evidence for epitaxial silicene on diboride thin films, \PRL{108}{245501}{2012}.
\bibitem{meng13} L. Meng, Y. Wang, L. Zhang, S. Du, R. Wu, L. Li, Y. Zhang, G. Li, H. Zhou, W. A. Hofer, and H.-J. Gao, Buckled silicene formation on Ir(111), \NL{13}{685}{2013}.
\bibitem{linn13} C.-L. Lin, R. Arafune, K. Kawahara, M. Kanno, N. Tsukahara, E. Minamitani, Y. Kim, M. Kawai, and N. Takagi, Substrate-induced symmetry breaking in silicene, \PRL{110}{076801}{2013}.
\bibitem{vogt12} P. Vogt, P. De Padova, C. Quaresima, J. Avila, E. Frantzeskakis, M. C. Asensio, A. Resta, B. Ealet, and G. Le Lay, Silicene: compelling experimental evidence for graphenelike two-dimensional silicon, \PRL{108}{155501}{2012}.
\bibitem{majz13} Z. Majzik, M. Rachid Tchalala, M. $\check{\rm S}$vec, P. Hapala, H. Enriquez, A. Kara, A. J. Mayne, G. Dujardin, P. Jel{\'{i}}ınek, and H. Oughaddou, Combined AFM and STM measurements of a silicene sheet grown on the Ag(111) surface, \JPCM{25}{225301}{2013}.
\bibitem{liuu14} Z.-L. Liu, M.-X. Wang, J.-P. Xu, J.-F. Ge, G. L. Lay, P. Vogt, D. Qian, C.-L. Gao, C. Liu, and J.-F. Jia, Various atomic structures of monolayer silicene fabricated on Ag(111), \NJP{16}{075006}{2014}.
\bibitem{chen12} L. Chen, C.-C. Liu, B. Feng, X. He, P. Cheng, Z. Ding, S. Meng, Y. Yao, and K.Wu, Evidence for dirac fermions in a honeycomb lattice based on silicon, \PRL{109}{056804}{2012}.
\bibitem{chen13} L. Chen, H. Li, B. Feng, Z. Ding, J. Qiu, P. Cheng, K. Wu, and S. Meng, Spontaneous symmetry breaking and dynamic phase transition in monolayer silicene, \PRL{110}{085504}{2013}.
\bibitem{chen13'} L. Chen, B. Feng, and K. Wu, Observation of a possible superconducting gap in silicene on Ag(111) surface surface, \APL{102}{081602}{2013}.
\bibitem{linn12} C.-L. Lin, R. Arafune, K. Kawahara, N. Tsukahara, E. Minamitani, Y. Kim, N. Takagi, and M. Kawai, Structure of silicene grown on Ag(111), \APE{5}{045802}{2012}.
\bibitem{john14} N. W. Johnson, P. Vogt, A. Resta, P. D. Padova, I. Perez, D. Muir, E. Z. Kurmaev, G. L. Lay, and A. Moewes, The metallic nature of epitaxial silicene monolayers on Ag(111), \AFM{24}{5253}{2014}.
\bibitem{tcha14} M. R. Tchalala, H. Enriquez, H. Yildirim, A. Kara, A. J. Mayne, G. Dujardin, M. A. Ali, and H. Oughaddou, Atomic and electronic structures of the ($\sqrt{13}$$\times$$\sqrt{13}$)R13.9$^{\circ}$ of silicene sheet on Ag(111), \ASS{303}{61}{2014}.
\bibitem{araf13} R. Arafune, C.-L. Lin, K. Kawahara, N. Tsukahara, E. Minamitani, Y. Kim, N. Takagi, and M. Kawai, Structural transition of silicene on Ag(111), \SSA{608}{297}{2013}.
\bibitem{pado13} P. De Padova, P. Vogt, A. Resta, J. Avila, I. Razado-Colambo, C. Quaresima, C. Ottaviani, B. Olivieri, T. Bruhn, T. Hirahara, T. Shirai, S. Hasegawa, M. C. Asensio, and G. Le Lay, Evidence of Dirac fermions in multilayer silicene, \APL{102}{163106}{2013}.
\bibitem{rest13} A. Resta, T. Leoni, C. Barth, A. Ranguis, C. Becker, T. Bruhn, P. Vogt, and G. Le Lay, Atomic structures of silicene layers grown on Ag(111): scanning tunneling microscopy and noncontact atomic force microscopy observations, \SR{3}{2399}{2013}.
\bibitem{guo14} Z. X. Guo and A. Oshiyama, Structural tristability and deep Dirac states in bilayer silicene on Ag(111) surfaces, \PRB{89}{155418}{2014}.
\bibitem{shir14} T. Shirai, T. Shirasawa, T. Hirahara, N. Fukui, T. Takahashi, and S. Hasegawa, Structure determination of multilayer silicene grown on Ag(111) films by electron diffraction: evidence for Ag segregation at the surface, \PRB{89}{241403}{2014}.
\bibitem{aiza99} H. Aizawa, M. Tsukada, N. Sato, S. Hasegawa, Asymmetric structure of the Si(111)-$\sqrt{3}$$\times$$\sqrt{3}$-Ag surface, \SSL{429}{L509}{1999}.
\bibitem{mann14} A. J. Mannix, B. Kiraly, B. L. Fisher, M. C. Hersam, and N. P. Guisinger, Silicon growth at the two-dimensional limit on Ag(111), \ACS{8}{7538}{2014}.
\bibitem{kres96} G. Kresse and J. Furthm{\"{u}}ller, Efficient iterative schemes for $ab$ $initio$ total-energy calculations using a plane-wave basis set, \PRB{54}{11169}{1996}.
\bibitem{perd96} J. P. Perdew, K. Burke, and M. Ernzerhof, Generalized gradient approximation made simple, \PRL{77}{3865}{1996}.
\bibitem{bloc94} P. E. Bl{\"{o}}chl, Projector augmented-wave method, \PRB{50}{17953}{1994}.
\bibitem{kres99} G. Kresse and D. Joubert, From ultrasoft pseudopotentials to the projector augmented-wave method, \PRB{59}{1758}{1999}.
\bibitem{guo15} Z. X. Guo and A. Oshiyama, Crossover between silicene and ultra-thin Si atomic layers on Ag, \NJP{17}{045028}{2015}.
\bibitem{kitt05} C. Kittel, Introduction to Solid State Physics, 8th ed. (Wiley, New York, 2005).
\bibitem{jons98} H. J{\'{o}}nsson, G. Mills, and K. W. Jacobsen, in Classical and Quantum Dynamics in Condensed Phase Simulations, edited by B. J. Berne, G. Ciccotti, and D. F. Coker (World Scientific, Singapore, 1998).
\bibitem{cord08} B. Cordero, V. G{\'{o}}mez, A. E. Platero-Prats, M. Rev{\'{e}}s, J. Echeverr{\'{i}}a, E. Cremades, F. Barrag{\'{a}}n, and S. Alvarez, Covalent radii revisited, \DT{21}{2832}{2008}.
\bibitem{vogt14} P. Vogt, P. Capiod, M. Berthe, A. Resta, P. De Padova, T. Bruhn, G. Le Lay, and B. Grandidier, Synthesis and electrical conductivity of multilayer silicene, \APL{104}{021602}{2014}.
\bibitem{chen15} J. Chen, Y. Du, Z. Li, W. Li, B. Feng, J. Qiu, P. Cheng, S. X. Dou, L. Chen, and K. Wu, Delocalized Surface State in Epitaxial Si(111) Film with Spontaneous $\sqrt{3}$$\times$$\sqrt{3}$ Superstructure, \SR{5}{13590}{2015}.
\end{thebibliography}
\end{document}